\documentclass{article}
\usepackage{graphicx} % Required for inserting images
\usepackage{hep-title}
\title{Are Parton Showers in a Quark-Gluon Plasma Strongly Coupled? A Theorist's Test}
\preprint{CERN-TH-2024-163}
\author[a]{Peter Arnold}
\author[b,c,a]{Omar Elgedawy}
\author[d,e]{Shahin Iqbal*}

\affiliation[a]{Department of Physics, University of Virginia,
 P.O. Box 400714, Charlottesville, VA 22904, U.S.A.}

\affiliation[b]{Key Laboratory of Atomic and Subatomic Structure and Quantum Control (MOE), Guangdong Basic Research Center of Excellence for Structure and Fundamental Interactions of Matter, Institute of
Quantum Matter, South China Normal University, Guangzhou 510006, China}

\affiliation[c]{Guangdong-Hong Kong Joint Laboratory of Quantum Matter,  Guangdong Provincial Key Laboratory of Nuclear Science,  Southern Nuclear Science Computing Center, South China Normal University,
Guangzhou 510006, China}

\affiliation[d]{National Centre for Physics, QAU Campus, Shahdra Valley Road, Islamabad, 45320 Pakistan}

\affiliation[e]{Theoretical Physics Department, CERN, CH-1211 Geneva 23, Switzerland}
\date{January 2025}

\begin{document}

\maketitle
\abstract{
We study whether in-medium showers of high-energy quarks and gluons can be treated as a sequence of individual splitting processes or whether there is significant quantum overlap between where one splitting ends and the next begins. Accounting for the Landau-Pomeranchuk-Migdal (LPM) effect, we calculate such overlap effects to leading order in high-energy $\alpha_s(\mu)$ for the simplest theoretical situation. We investigate a measure of overlap effects that is independent of physics that can be absorbed into an effective value $\hat{q}_{eff}$ of the jet-quenching parameter $\hat{q}$.
}
\pagebreak

\section{Introduction}

High energy particles passing through a medium primarily lose energy through splitting processes of bremmstrahlung and pair-production. At high enough energies, the quantum-mechanical duration of the splitting process, called the "formation time", exceeds the mean free time between elastic collisions in the medium, leading to a significant reduction in the splitting rate due to the Landau-Pomeranchuk-Migdal (LPM) effect \cite{LP1,LP2,Migdal}. QCD generalization of the LPM effect was developed in the 1990s \cite{BDMPS1,BDMPS2,BDMPS3,Zakharov1,Zakharov2}. A long standing problem in studying in-medium evolution of high energy QCD showers in relativistic heavy ion collisions has been calculating corrections to the LPM effect for cases where consecutive splittings happen close enough that their formation times overlap. That is, whether consecutive splittings in a high energy shower can be treated as being probabilistically independent, or if there are significant quantum interference effects that must be accounted for.

Beginning several years ago, together with collaborators, we have worked to answer this question, culminating with our complete result for an all gluon, i.e. $N_f=0$ in-medium shower \cite{finale,finale2}. Here $N_f$ refers to the number of electron or quark flavors. We found that the effect of overlapping formation times for an all gluon QCD shower was a nearly negligible $O(1\%)$ effect. This is in stark contrast to the $O(100\%)$ effect that we found for the case of Large-$N_f$ QED \cite{qedNfstop,qedNfenergy,qedNf} earlier. It raises the important question: What exactly makes overlap corrections in these two cases so radically different?  Large-$N_f$ QED is, of course different from $N_f=0$ QCD. But is the large difference in the size of overlap effects due to ignoring fermions in refs. \cite{finale,finale2}, or is it related more fundamentally to the structure of QCD? To answer this question, we consider now the other extreme, i.e. QCD in the large-$N_f$ limit!

Following our previous work, we will continue to make certain simplifying assumptions: i) We will formally assume the medium to be infinitely large, homogeneous and static, ii) we will work in the multiple scattering (also called the $\hat{q}$) approximation, and iii) we will work in the large-$N_c$ limit of QCD, $N_c$ being the number of colors. Specifically, we will assume $N_f\gg N_c\gg 1$.   

\section{Overview of the calculation}
To qualitatively understand the LPM effect, consider as example the case of an electron moving through a medium. The electron scatters around, and eventually radiates a bremsstrahlung photon. The photon cannot, obviously resolve details smaller than its wavelength. For any external observer, it then creates an uncertainty about the exact time and location of the splitting process. The size of this uncertainty is often referred to as the "formation time" or "formation length". In situations where the wavelength, and hence the formation time of the radiated photon becomes larger than the mean free time between collisions in the medium, a bremsstrahlung resulting from a single scattering in the medium becomes indistinguishable from one resulting from many small angle scatterings.  Therefore, the observed splitting rate ends up being smaller than what one would naively have expected. 

It is important to note that the LPM effect in QCD is qualitatively different from that in QED. Unlike the case of photon bremsstrahlung, the LPM suppression is smaller for softer gluons. This is because, unlike electrically neutral photons, gluons carry color and interact strongly with the medium. A soft gluon is therefore easier to deflect, and gets separated from its parent quickly, which reduces its formation time and consequently reducing the resulting LPM suppression. 

\subsection{Weakly-vs. strongly coupled showers}
Once we have accounted for the leading order LPM effect, can we then treat consecutive splitting processes in an idealized Monte-Carlo simulation as being probabilistically independent? In effect, rolling a classical dice at each time step with the probability of radiation weighted by the LPM splitting rate? The answer to this question depends on the value of the QCD coupling associated with the high energy splitting. For the case of democratic splittings, for which the formation times are the largest, the time between splittings is parametrically $t_{rad}\sim \frac{t_{form}}{\alpha_s}$. Hence, when $\alpha_s$ is small, the shower will be made up of well-separated splitting and the chance of overlap will be negligible and the shower may be approximated as a series of independent splitting processes. However, the value of $\alpha_s$ reached in real life heavy ion collisions is only moderately small. Besides previous authors have shown that corrections from soft bremsstrahlung give large double-logarithmic enhancements, which can be absorbed into an effective value of the $\hat{q}$ \cite{Wu,Blaizot,Iancu}. So now we might ask a more refined question:\textit{ How big are overlap effects that cannot be absorbed into an effective value of $\hat{q}$?} 

\subsection{A $\hat{q}-$independent measure of overlap effects}

Consider a shower made up entirely of democratic splittings. The shower will completely stop in the medium in a distance $l_{stop}\sim \alpha^{-1}\sqrt{E_0/\hat{q}}$. As a theorist's thought experiment, one might imagine measuring the distribution $\epsilon(z)$ of energy deposited by the shower as it moves through the medium. Then, $l_{stop}$ will be the first moment of the distribution, i.e.,

\begin{equation}
	l_{stop}=\langle z \rangle=E_0^{-1}\int_{z} z \epsilon(z).
\end{equation}

The width $\sigma=\sqrt{\langle z^{2}\rangle-\langle z \rangle^{2}}$ is parametrically of the same order as $l_{stop}$, i.e $\sigma\sim \alpha^{-1}\sqrt{E_0/\hat{q}}$ and any ratio of these quantities will be independent of $\hat{q}$. One such quantity is $\frac{\sigma}{l_{stop}}$ for which one can calculate corrections due to overlapping formation times. That is, we may formally write 
\begin{equation}
	\frac{\sigma}{l_{stop}}=\left(\frac{\sigma}{l_{stop}}\right)_{LO}\left(1+\chi\alpha + O(\alpha^2)+...\right)
\end{equation}
and calculate the value of $\chi$ as a measure of the effect of overlapping formation times that cannot be absorbed into $\hat{q}$. % In our previous work, we calculated this correction for the case of large$-N_f$ QED \cite{qedNf,qedNfstop,qedNfenergy}, where we found that the correction due to overlapping formation times is substantial and $O(100\%)\times N_f \alpha_{QED}$. However, in a similar study for a pure gluon ($N_f=0$) shower in large$-N_c$ QCD we found that the correction is tiny $O(1\%)\times N_c \alpha_s$ \cite{qcdNf}. To understand why the effect of overlapping formation times is so radically different between QED and QCD, and whether it is related to quarks, we now consider the case of a quark and gluon shower for QCD in the limit $N_f\gg N_c\gg 1$. 
\subsection{QCD in $N_f\gg N_c\gg 1$ and other approximations}

\begin{figure}[h]
	\centering
	\includegraphics[width=0.3\textwidth]{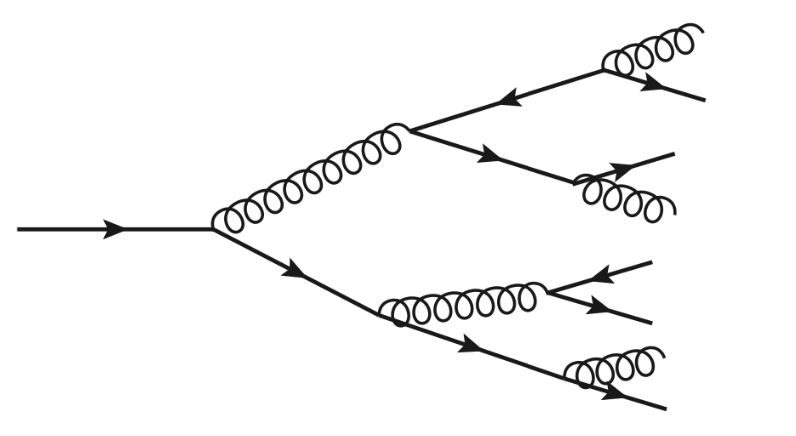}
	\caption{A typical shower in $N_f\gg N_c\gg 1$ limit of QCD. }
	\label{Fig3}
\end{figure}

We will make certain simplifying assumptions. Specifically, we will work in the $N_f\gg N_c\gg 1$ limit of QCD, in which a typical shower will be entirely made up of $q\to q g$ and $g\to q \bar{q}$ splitting vertices as shown in Fig. \ref{Fig3}. We will also assume the medium to be static, homogenous and large compared to the entire shower.  We will ignore vacuum as well as medium-induced masses of all high energy particles. We will assume that the high energy particle initiating the shower can be approximated to be on-shell and finally, we will integrate over all transverse degrees of freedom. 
\subsection{Brief overview of the calculation}
\begin{figure}[h]
\begin{minipage}[c]{0.4\linewidth}
\centering
	\includegraphics[width=1.2\textwidth]{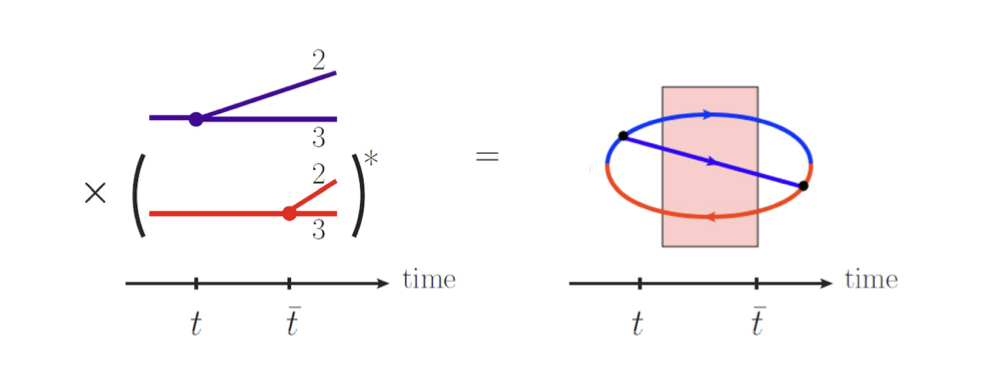}
	\caption{An interference contribution to leading order LPM effect. }
	\label{Fig4}
	\end{minipage}\hfill
\begin{minipage}[c]{0.4\linewidth}
	\centering
	\includegraphics[width=\textwidth]{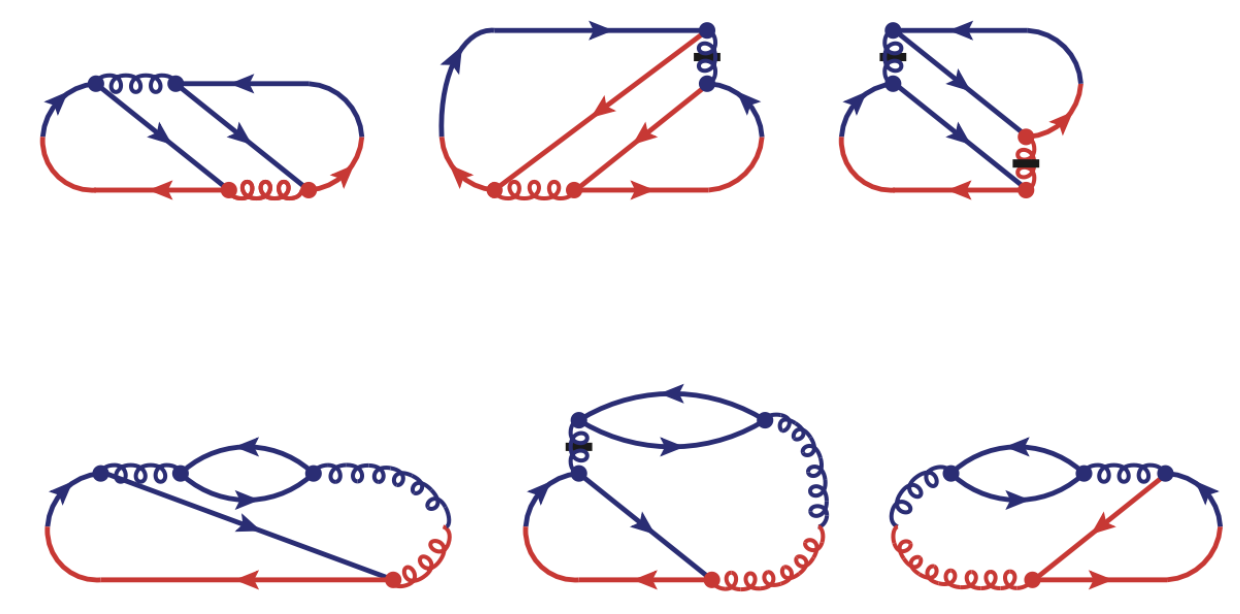}
	\caption{A small subset of real and virtual, NLO interference contributions that we calculate.}
	\label{Fig5}
	\end{minipage}
\end{figure}

In terms of feynman diagrams, the LPM effect arises from quantum interference of splittings amplitudes from splittings at slightly different times as shown in Fig. \ref{Fig4}. For ease of notation, we have chosen not to show medium interactions, and have only drawn the high energy in-medium particles, depicting them as solid straight lines. The upper blue lines represent the splitting at time $t$, while the lower red represents the same splitting process, except at a slightly different time $\bar{t}$. For calculations like these it turns out to be useful to draw these interference contributions as a single process, as shown on the right in Fig. \ref{Fig4}. In this way, the leading order LPM interference contribution can be interpreted as a 3-particle in-medium evolution sandwiched between splitting matrix elements at times $t$ and $\bar{t}$. The in-medium evolution between $t$ and $\bar{t}$ can be reduced to an effective, non-relativistic quantum mechanics problem governed by a non-Hermitian Hamiltonian. The splitting matrix elements are related to the usual QCD Feynman rules. 

The next-to-leading order, overlapping formation time contributions are calculated in similar, although much more complicated way. A sample of such diagrams is shown in Fig. \ref{Fig5}.
\section{Results}

Surprisingly, we find that $\hat{q}$-insensitive correction from overlap effects $\chi \alpha \sim 0.005$, which again is drastically different from our results in QED. Clearly, the small result in QCD does not appear to be a peculiarity of purely gluonic showers. Adding many flavors of quarks does not qualitatively change the size of overlap effects in QCD. 

\begin{figure}
	\centering
	\includegraphics[width=0.6\textwidth]{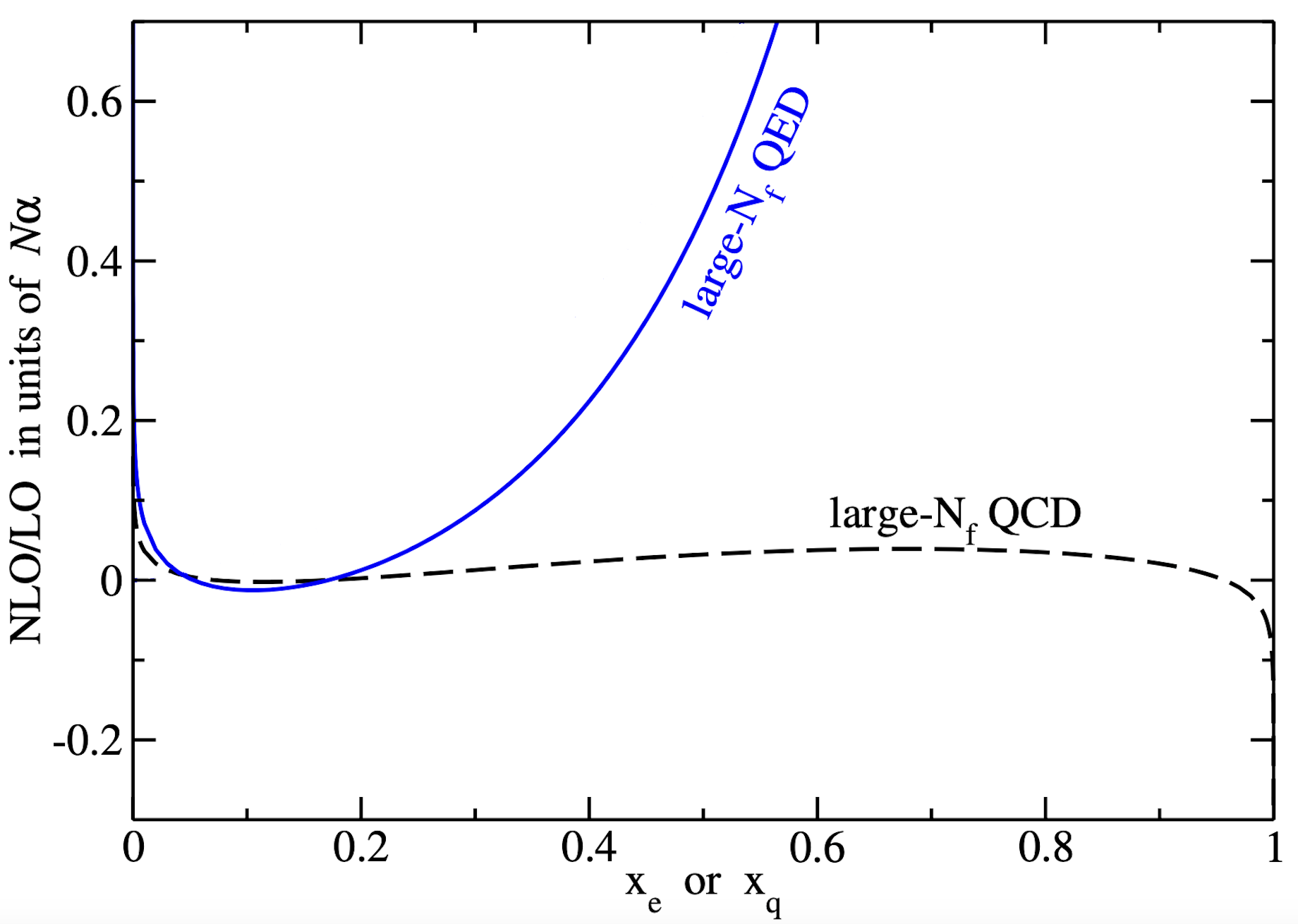}
	\caption{Ratio of NLO to LO contributions to $q\to q g$ and $e\to e \gamma$.}
	\label{Fig6}
\end{figure}

To understand why the result is so different between QED and QCD, consider the behavior of LO and NLO splitting rates shown in Fig. \ref{Fig6}. Here we plot the ratio of NLO to LO contributions to the $q\to qg$ splitting rate for both QED and QCD in the large$-N_f$ limits. As is clear from the plot, the NLO contribution grows rapidly as the energy fraction carried by the final state electron/quark  $x_e\to 1$.  In other words, the NLO correction is large when the intermediate photon becomes soft. However, we do not see any such behavior for the intermediate gluon in the case of QCD. 

We will present a brief, qualitative argument explaining why overlap effects become significant for soft intermediate photons but not for soft intermediate gluons. For soft bremsstrahlung, formation times and hence splitting rates behave very differently in QED and QCD. For QED, 
\begin{equation}
	t_{form}\sim \sqrt{\frac{\hat{q}E}{x_\gamma }}
\end{equation}
while for QCD, 
\begin{equation}
	t_{form}\sim \sqrt{\hat{q}E x_\gamma } .
\end{equation}
Hence, at leading order, the LPM suppression grows as the radiated photon becomes soft. A softer photon has a larger wavelength, hence increasing the uncertainty about the exact time and place of the splitting process for any external observer, and hence has larger formation time, and larger LPM suppression. Now if a subsequent overlapping $\gamma\to e\bar{e}$ splitting happens (which is the only possible scenario in large$-N_f$ QED), the resulting electron-positron pair, unlike the photon, carries electric charge and directly interacts with QED plasma, receiving transverse momentum kicks from the medium and getting deflected. Hence any subsequent, overlapping pair production then significantly disrupts the collinearity  of the original $e\to e\gamma $ bremsstrahlung. Therefore, the correction from overlapping $e\to e \gamma \to e\bar{e}$ is substantial in QED. 
In QCD on the other hand, quarks and gluons both carry color and interact strongly with the QCD medium. A subsequent overlapping $g\to q\bar{q}$ pair-production does not significantly affect the collinearity of the original $q\to q g$ bremsstrahlung.  Consequently the correction from overlapping $q\to q g\to q Q\bar{Q}$ is therefore small. We will refer the readers to our recent paper \cite{qcdNf} for more detailed discussion of this result. 

\section{Conclusion}
Our results demonstrate that the small size of overlap effects in pure gluon ($N_f=0$)  QCD were not a numerical accident, \textit{soft photons are affected much more significantly by a subsequent pair production than soft gluons are.} Leaving the question of overlap effects for the case of $N_c\sim N_f$ for future work, we can conclude for now that overlap effects that cannot be absorbed into an effective value of $\hat{q}$ are small for both $N_f=0$ and $N_f\gg 1$ limits of QCD.


\begin{thebibliography}{99}

\bibitem{LP1}
  L.~D.~Landau and I.~Pomeranchuk,
  ``Limits of applicability of the theory of bremsstrahlung electrons and
  pair production at high-energies,''
  Dokl.\ Akad.\ Nauk Ser.\ Fiz.\  {\bf 92} (1953) 535.

\bibitem{LP2}
  L.~D.~Landau and I.~Pomeranchuk,
  ``Electron cascade process at very high energies,''
  Dokl.\ Akad.\ Nauk Ser.\ Fiz.\  {\bf 92} (1953) 735.

\bibitem{Migdal}
  A.~B.~Migdal,
  ``Bremsstrahlung and pair production in condensed media at high-energies,''
   Phys.\ Rev.\  {\bf 103}, 1811 (1956);


\bibitem{BDMPS1}
  R.~Baier, Y.~L.~Dokshitzer, A.~H.~Mueller, S.~Peigne and D.~Schiff,
  ``The Landau-Pomeranchuk-Migdal effect in QED,''
  Nucl.\ Phys.\  B {\bf 478}, 577 (1996)
  [arXiv:hep-ph/9604327];

\bibitem{BDMPS2}
  R.~Baier, Y.~L.~Dokshitzer, A.~H.~Mueller, S.~Peigne and D.~Schiff,
  ``Radiative energy loss of high-energy quarks and gluons in a
    finite volume quark-gluon plasma,''
  Nucl.\ Phys.\  B {\bf 483}, 291 (1997) [arXiv:hep-ph/9607355].
  %%CITATION = NUPHA,B483,291;%%

\bibitem{BDMPS3}
  R.~Baier, Y.~L.~Dokshitzer, A.~H.~Mueller, S.~Peigne and D.~Schiff,
  ``Radiative energy loss and $p_\perp$-broadening of high energy partons in
    nuclei,''
  Nucl.\ Phys.\  B {\bf 484} (1997)
  [arXiv:hep-ph/9608322].
  %%CITATION = NUPHA,B484,265;%%

\bibitem{Zakharov1}
 B.~G.~Zakharov,
 ``Fully quantum treatment of the Landau-Pomeranchuk-Migdal effect in
   QED and QCD,''
 JETP Lett.\  {\bf 63}, 952 (1996)
 [Pis'ma Zh.\ \'Eksp.\ Teor.\ Fiz.\  {\bf 63}, 906 (1996)]
 [arXiv:hep-ph/9607440].

\bibitem{Zakharov2}
 B.~G.~Zakharov,
 ``Radiative energy loss of high-energy quarks in finite size nuclear matter
   and quark-gluon plasma,''
 JETP Lett.\  {\bf 65}, 615 (1997)
 [Pis'ma Zh.\ \'Eksp.\ Teor.\ Fiz.\  {\bf 65}, 585 (1997)]
 [arXiv:hep-ph/9704255].
 %%CITATION = JTPLA,63,952.%%

\bibitem{Blaizot}
  J.~P.~Blaizot and Y.~Mehtar-Tani,
  ``Renormalization of the jet-quenching parameter,''
  Nucl.\ Phys.\ A {\bf 929}, 202 (2014)
  [arXiv:1403.2323 [hep-ph]].
  %%CITATION = ARXIV:1403.2323;%%\bibitem{LP1}
  L.~D.~Landau and I.~Pomeranchuk,
  ``Limits of applicability of the theory of bremsstrahlung electrons and
  pair production at high-energies,''
  Dokl.\ Akad.\ Nauk Ser.\ Fiz.\  {\bf 92} (1953) 535.


\bibitem{Iancu}
  E.~Iancu,
  ``The non-linear evolution of jet quenching,''
  JHEP \textbf{10}, 95 (2014)
  [arXiv:1403.1996 [hep-ph]].
  %%CITATION = ARXIV:1403.1996;%%

\bibitem{Wu}
  B.~Wu,
  ``Radiative energy loss and radiative $p_{\bot}$-broadening of
    high-energy partons in QCD matter,''
  JHEP \textbf{12}, 081 (2014)
  [arXiv:1408.5459 [hep-ph]].
  %%CITATION = ARXIV:1408.5459;%%

\bibitem{finale}
  P.~Arnold, O.~Elgedawy and S.~Iqbal,
  ``Are gluon showers inside a quark-gluon plasma strongly coupled?
    a theorist's test,''
  Phys. Rev. Lett. \textbf{131}, no.16, 162302 (2023)
  %doi:10.1103/PhysRevLett.131.162302
  [arXiv:2212.08086 [hep-ph]].

\bibitem{finale2}
  P.~Arnold, O.~Elgedawy and S.~Iqbal,
  ``The LPM effect in sequential bremsstrahlung: gluon shower development,''
  Phys. Rev. D \textbf{108}, no.7, 074015 (2023)
  %doi:10.1103/PhysRevD.108.074015
  [arXiv:2302.10215 [hep-ph]].



\bibitem{qedNfstop}
  P.~Arnold, S.~Iqbal and T.~Rase,
  ``Strong- vs. weak-coupling pictures of jet quenching: a dry run using QED,''
  JHEP \textbf{05}, 004 (2019)
  %doi:10.1007/JHEP05(2019)004
  [arXiv:1810.06578 [hep-ph]].

\bibitem{qedNf}
  P.~Arnold and S.~Iqbal,
  ``In-medium loop corrections and longitudinally polarized gauge bosons
    in high-energy showers,''
  JHEP \textbf{12}, 120 (2018)
  %doi:10.1007/JHEP12(2018)120
  [erratum: JHEP \textbf{12}, 098 (2023)]
  [arXiv:1806.08796 [hep-ph]].

\bibitem{qedNfenergy}
  P.~Arnold, O.~Elgedawy and S.~Iqbal,
  ``Strong vs.\ weakly coupled in-medium showers: energy stopping in
    large-$N_f$ QED,''
  arXiv:2404.19008 [hep-ph].


\bibitem{qcdNf}
P. Arnold, and O. Elgedawy, and S. Iqbal,
   "Are in-medium quark-gluon showers strongly coupled? Results in the large-$N_f$ limit",
   [arXiv: 2408.07129 [hep-ph]].
   


\end{thebibliography}
\end {document}